\documentclass[conference,10pt,letterpaper]{IEEEtran}
\usepackage{cite}
\usepackage{amsmath,amssymb,amsfonts}
\usepackage{algorithmic}
\usepackage{graphicx}
\usepackage{textcomp}
\usepackage{xcolor}
\usepackage{caption}
\def\BibTeX{{\rm B\kern-.05em{\sc i\kern-.025em b}\kern-.08em
    T\kern-.1667em\lower.7ex\hbox{E}\kern-.125emX}}

\renewcommand{\thesection}{\textbf{\Roman{section}}}

\usepackage{hyperref}

\begin{document}

\title{\fontsize{24}{24}\selectfont{Convolutions with Radio-Frequency Spin-Diodes}}
\author{\fontsize{11}{11}\selectfont E. Plouet, H. Singh, P. Sethi, F.A. Mizrahi, D. Sanz-Hern\'{a}ndez and J. Grollier\\
\fontsize{10}{12}\selectfont Laboratoire Albert Fert, CNRS, Thales, Universit\'e Paris-Saclay, 91767 Palaiseau, France\\ email: dedalo.sanz@cnrs-thales.fr, julie.grollier@cnrs-thales.fr\\ \vspace{-1ex}
}

\maketitle
\noindent \textbf{\textit{Abstract}}---The classification of radio-frequency (RF) signals is crucial for applications in robotics, traffic control, and medical devices. Spintronic devices, which respond to RF signals via ferromagnetic resonance, offer a promising solution. Recent studies have shown that a neural network of nanoscale magnetic tunnel junctions can classify RF signals without digitization. However, the complexity of these junctions poses challenges for rapid scaling. In this work, we demonstrate that simple spintronic devices, known as metallic spin-diodes, can effectively perform RF classification. These devices consist of NiFe/Pt bilayers and can implement weighted sums of RF inputs. We experimentally show that chains of four spin-diodes can execute 2x2 pixel filters, achieving high-quality convolutions on the Fashion-MNIST dataset. Integrating the hardware spin-diodes in a software network, we achieve a top-1 accuracy of 88 \% on the first 100 images, compared to 88.4 \% for full software with noise, and 90 \% without noise.\vspace{-0.5ex}

\section{\textbf{Introduction}}

Fully spintronic neuronal networks using nanoscale magnetic tunnel junctions (MTJs) can classify non-linearly separable radio-frequency (RF) inputs without pre-processing or digitization \cite{ross2023multilayer}. Such a system is predicted to significantly improve energy efficiency, consuming about 10 fJ per synapse and 100 fJ per neuron with MTJs miniaturized to 20 nm—over 100 times more efficient than GPUs. The frequency multiplexing strategy used for neuron-to-synapse communication supports high connectivity, up to 500 synapses per neuron, and is free of sneak paths. However, current hardware implementations are limited to two synapses per neuron due to the complexity of MTJ stacks. To address these challenges, simpler materials can be used, maintaining the principles of RF interconnects and frequency multiplexing. In this work, we demonstrate that two-layer metallic spin-diodes \cite{finocchio2021perspectives} can perform high-quality weighted sums, enabling convolutional filters on the Fashion-MNIST dataset \cite{xiao2017fashion}.

\section{\textbf{Metallic spin-diode synapses}}

The samples studied, shown in Fig. 1(a), consist of a 5nm nickel iron bilayer capped with 5nm platinum on a high-resistance silicon wafer. Fig. 1(b) and (c) display a typical device contacted with gold electrodes, measuring 5 by 10 micrometers. When an RF current is injected into these stripes, they generate a rectified DC voltage $V_{DC}$ due to the mixing of the input with the anisotropic magneto-resistance (AMR) response, which can be calculated analytically \cite{azevedo2011spin}.

Fig. 2(a) shows the analytically calculated variations of $V_{DC}$ as a function of the applied magnetic field amplitude $H$ along the z-axis for a fixed frequency RF current. The applied magnetic field monotonically varies the device's resonant frequency, $f_{res} = f_{res}(H)$, from $\approx$ 1 to 4 GHz here. The voltage is positive when $f_{res}(H) < f_{RF}$  and negative when $f_{res}(H) > f_{RF}$. For small differences between $f_{res}(H)$ and $f_{RF}$, the voltage simplifies to: $V_{DC}\propto  
(f_{res}(H)-f_{RF}) sin(2\Phi_{0})P_{RF} \propto W P_{RF}$, with $P_{RF}$ the power of the injected current.

In our design (Fig. 1(c)), we set the angle $\Phi_{0}$ between the stripe and the applied magnetic field to 45° to maximize $V_{DC}$. Fig. 2(b) shows the calculated $V_{DC}$ as a function of $P_{RF}$ for different magnetic field values. The RF power of the injected current ($P_{RF}$) is the synapse input, and $V_{DC}$ is the output. The weight $W$, corresponding to the slope in Fig. 2(b), is tuned by the applied magnetic field.

To perform the weighted sum, we use three methods illustrated in Fig. 3. First, we electrically connect several devices to sum their DC voltages. Second, we send RF inputs ($P_{RFi}$) at different frequencies: $I_{RF}(tot) = \sum_i I_{RFi}$. Third, we apply different magnetic fields to each device, giving each a unique resonant frequency $f_{res}(Hi)$ close to $f_{RFi}$. This frequency multiplexing technique implements a weighted sum, with, in the linear regime,  
$V_{DC}\propto  \sum_i
(f_{res}(Hi)-f_{RFi}) sin(2\Phi_{0})P_{RFi} \propto \sum W_i P_{RFi}$.

Fig. 4(a) shows the experimental implementation of a spatial field gradient along the z direction with four permanent magnets. This gradient links the synaptic weight of each device to its position along the z-axis. Fig. 5 shows the field components $H_x$, $H_y$, and $H_z$ obtained by spatially scanning a 3D Hall probe in the field gradient with a motorized stage, achieving large variations in $H_z$ over +/- 10 mT along the z-axis while keeping small variations in other directions.

Fig. 6(a) displays the experimentally measured $V_{DC}$ response of a single device as a function of the z position in the field gradient. Fig. 6(b) shows the $V_{DC}$ vs. $P_{RF}$ curves for different z positions, qualitatively aligned with the analytical calculations in Fig. 2. The large background signal outside the linear operation region (highlighted in yellow) is problematic for devices in a chain due to overlaps.

We design differential RF synapses to cancel these backgrounds, leaving a strong response in the desired region only. As shown in Fig. 7, such synapse comprises two NiFe/Pt stripes connected in parallel at +45° and –45° to the magnetic field. This angle difference produces opposite $V_{DC}$ signals. 

Fig. 8(b) shows that when fabricated at the same z position, the overall $V_{DC}$ cancels out over a wide range of RF inputs. When slightly displaced by $\delta_z$ (40 micrometers in Fig. 8(a) and (c)), the signal response is strong within a narrow region, demonstrating frequency selectivity. The peak amplitude is proportional to $\delta_z$ and $P_{RF}$, making the synaptic weight controllable by the shift between devices (Fig. 8(d)).

To predict weight values before lithography, we perform a calibration step. We record and average the response of three identical 45° stripes for various $f_{RF}$ and $z$ positions, obtaining a reference $V_{ref}(f, z)$ (Fig. 9a). The differential pair's synaptic response is predicted by $V_{synapse}(f, z) = 0.5  \times (V_{ref} (f, z – \delta_z / 2 ) + V_{ref} (f, z + \delta_z /2 ))$. This response is shown in Fig. 9b for fixed $P_{RF}$ and $\delta_z$, with peak values decreasing with input frequency. This effect is accounted for by rescaling the injected power: $P_{RF}(cal)=P_{RF}\times (-40\times 3.1(GHz) +150) / 
(-40\times f_{RF}(GHz) +150)$.

For weighted sum operations, the device chains must have an impedance matched to 50 $\Omega$  RF sources for effective signal propagation. Various architectures are possible: series (Fig. 10(a)), parallel (Fig. 10(b)), or hybrid (Fig. 10(c)). We choose the hybrid configuration because the total chain resistance does not depend on the number of devices it contains, which will enable the implementation of larger kernels (of size > 9x9) in the future. Examples of fabricated devices with a target resistance of 50 $\Omega$ are shown in Fig. 11(a) and (b).

\section{\textbf{Spin-diode convolutions}}

We use three chains of four differential synapses to implement three 2x2 convolutional filters on the Fashion MNIST dataset, as illustrated in Fig. 12.The twelve weights are pre-calculated by training a software neural network, as shown in Fig. 13. The network consists of a convolutional layer with three 2x2 filters (stride and padding of 1), followed by max pooling (kernel size and stride of 2), ReLU neurons, a fully connected layer, and a softmax function for classification. During training, Gaussian noise is added to the outputs of the convolutional layer to ensure noise-resistant weights. The test accuracy reaches 87.63\% without noise and 86.34\% with 50\% noise, indicating a small accuracy loss of 1.3\%.

After training, we implement the learned weights of the first convolutional layer in hardware through lithography. The input frequencies are set to 1.75, 2.25, 2.75, and 3.25 GHz. Based on the interpolation of a single diode's resonance frequency vs. its position along the z-axis, synapses are spaced with 175, 216, and 245 micrometers to center their frequency peaks at these values. Fig. 14 shows the targeted weights, corresponding $\delta_z$ values, and measured weights after fabrication. Discrepancies arise from resolution limitations in photolithography leading to unwanted variations in  width and position of the diodes. Fig. 15 compares the predicted profile (red) from reference diode interpolation and the experimental profile (blue) of fabricated samples. Fig. 16 compares the experimental versus predicted operations for each chain.

To perform the convolution, each 2x2 region is flattened into a 4-pixel vector. Each pixel is converted to an RF signal, with its frequency encoding the pixel location and power encoding the pixel value (0 to 255). The four RF inputs are combined and sent to three chains of four synapses, each implementing a different convolutional filter. The output voltages are converted back to pixel values to form the three output images. To scan the entire input image, the kernels are slid with a stride of 2.

%The experimental accuracy on the classification of the first 100 images is 88\%, compared to 88.4\% for noisy software and 90\% without noise. This shows that our experimental network performs similarly to the noisy software model. The software accuracy with noise is calculated by passing each of the 100 input images 1000 times through the network with different random noise.

The experimental network reached an accuracy of 88\% on the first 100 images, close to the 88.4\% accuracy of the noisy software model, and slightly lower than the 90\% accuracy of the model without noise. The accuracy of the noisy software model was determined by passing each of the 100 input images through the network 1000 times with varying random noise.

The confusion matrices for the noisy software model and the experimental model (Fig. 17(a) and (b)) show similar misclassifications of pullovers, shirts, and coats, highlighting the comparable accuracy of both models for these relatively similar types of clothes. Fig. 18(a) and (b) show an example image after convolution in (a) software and (b) hardware. The difference (c) is close to zero, demonstrating the high quality of the hardware convolutions with RF metallic spin-diodes.

\section{\textbf{Conclusion}}

This work demonstrated the use of NiFe/Pt bilayer spin-diodes for high-quality RF signal classification. By employing chains of differential RF synapses, we implemented convolutional filters on the Fashion MNIST dataset, achieving a top-1 accuracy of 88\% on the first 100 images. This performance closely matches the 88.4\% accuracy of the noisy software model and 90\% without noise. 

This study underscores the promise of spintronic hardware for next-generation neural networks, offering significant energy efficiency and scalability advantages.  Future work will focus on non-volatile frequency control of diodes through magneto-ionic effects \cite{khademi2023large}, removing the need of field gradients, and unlocking on-chip training. Miniaturization  ofsynapses down to 10 nm, and their integration in large networks for more complex applications, will pave the way for efficient and scalable RF signal classification systems.

\section*{Acknowledgment}
This work received funding by the European Research
Council advanced grant GrenaDyn (reference: 101020684)\\

% Dummy (non-printed) citations to trigger automatic sorting and inclusion of the references.

%\nocite{fontnote}
%\nocite{fuller}
%\nocite{vidmar}
%\nocite{clarke}
%\nocite{reber}
%\nocite{yorozu}
%\nocite{alqueres}

% INSERT REFERENCE LIST
\bibliography{references}

\newpage
\onecolumn
\noindent
\begin{minipage}{0.53\textwidth}
    \includegraphics[width=\textwidth]{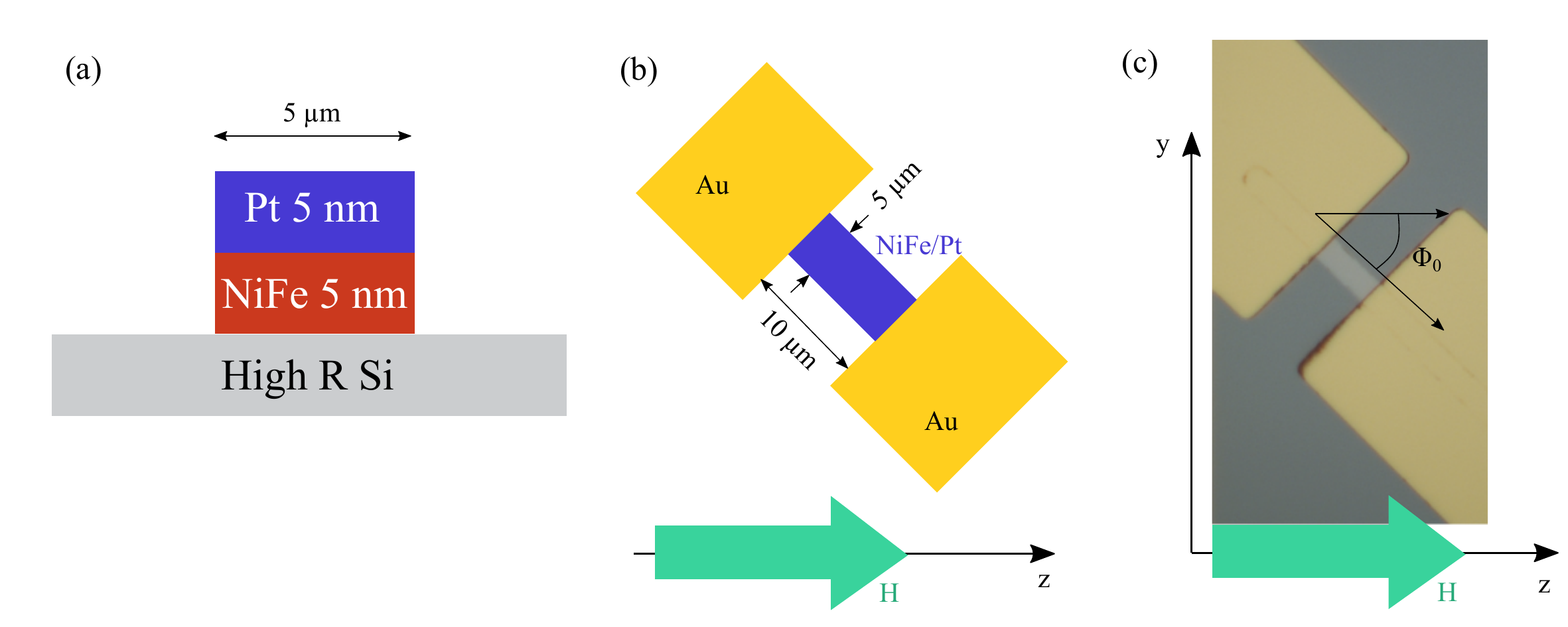}
    \captionof{figure}{Spin diode stack (a) and geometry (b)(c).}
    \label{fig:figure1}
\end{minipage}
\noindent
\begin{minipage}{0.43\textwidth}
    \includegraphics[width=\textwidth]{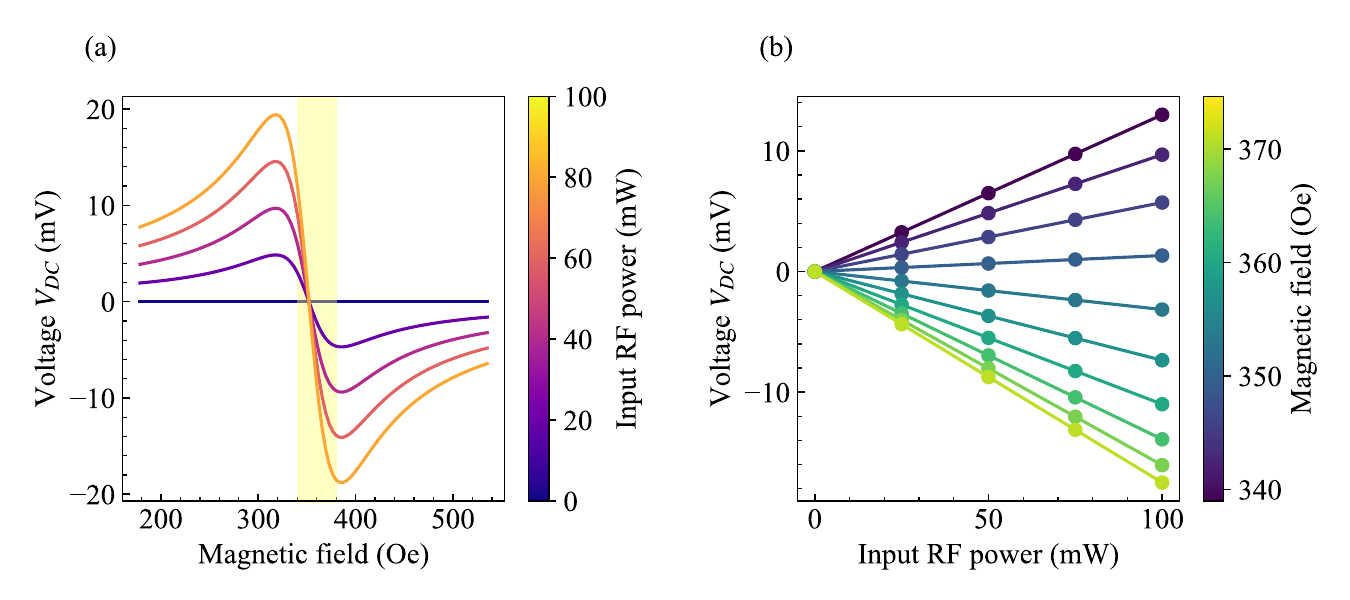}
    \captionof{figure}{(a) Theoretical spin diode voltage while varying the magnetic field, (b) synaptic behavior.}
    \label{fig:figure1}
\end{minipage}
\noindent
\begin{minipage}{0.5\textwidth}
    \includegraphics[width=\textwidth]{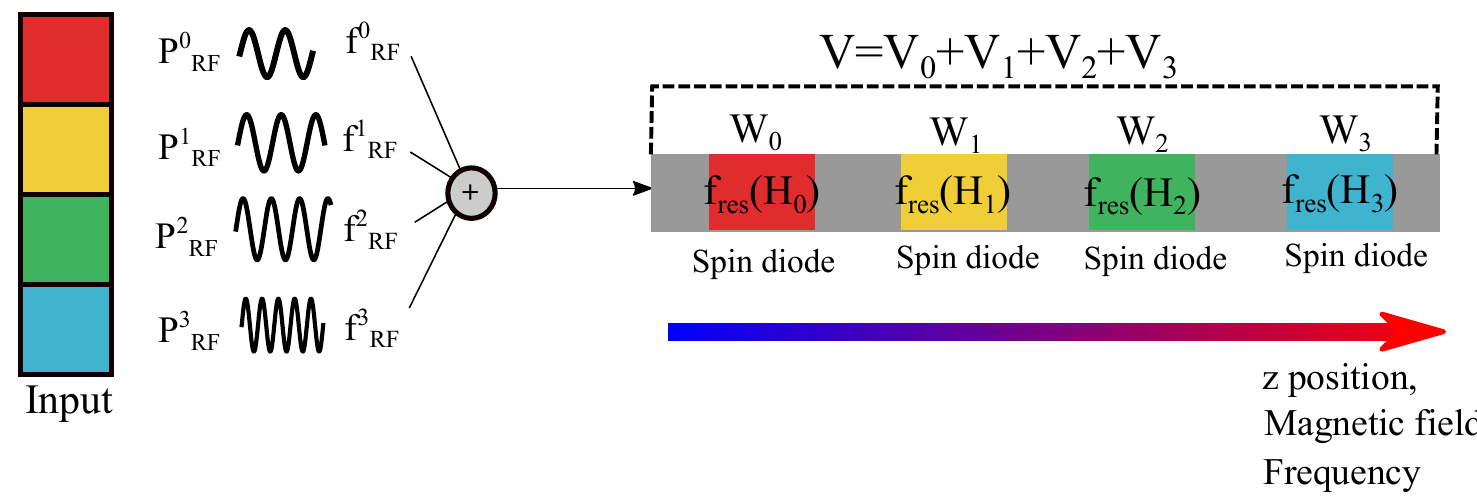}
    \captionof{figure}{Implementation of a spintronic weighted sum performed on frequency multiplexed inputs}
\end{minipage}
\noindent
\begin{minipage}{0.4\textwidth}
    \includegraphics[width=\textwidth]{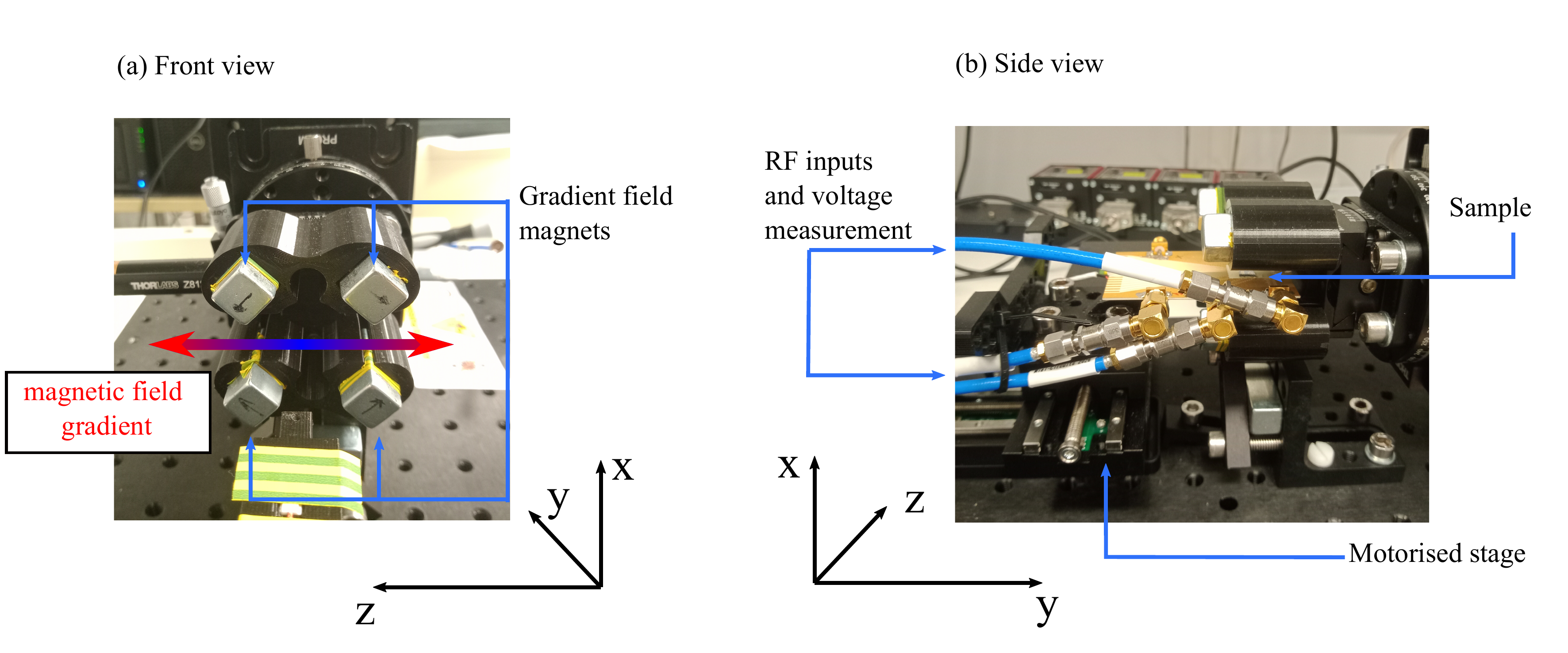}
    \captionof{figure}{Experimental measurement setup}
\end{minipage}
\noindent
\begin{minipage}{0.55\textwidth}
    \includegraphics[width=\textwidth]{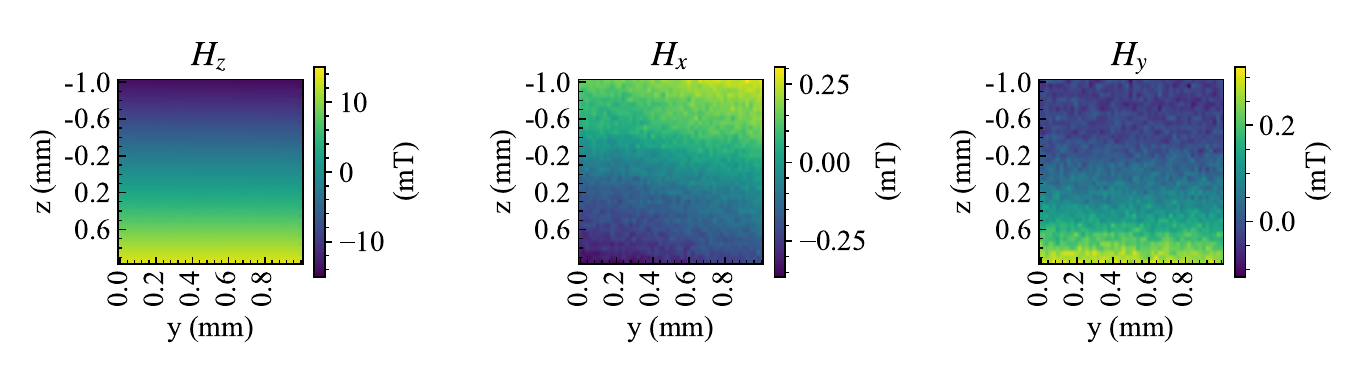}
    \captionof{figure}{Measurement of the magnetic field created with the gradient magnets}
\end{minipage}
\noindent
\begin{minipage}{0.43\textwidth}
    \includegraphics[width=\textwidth]{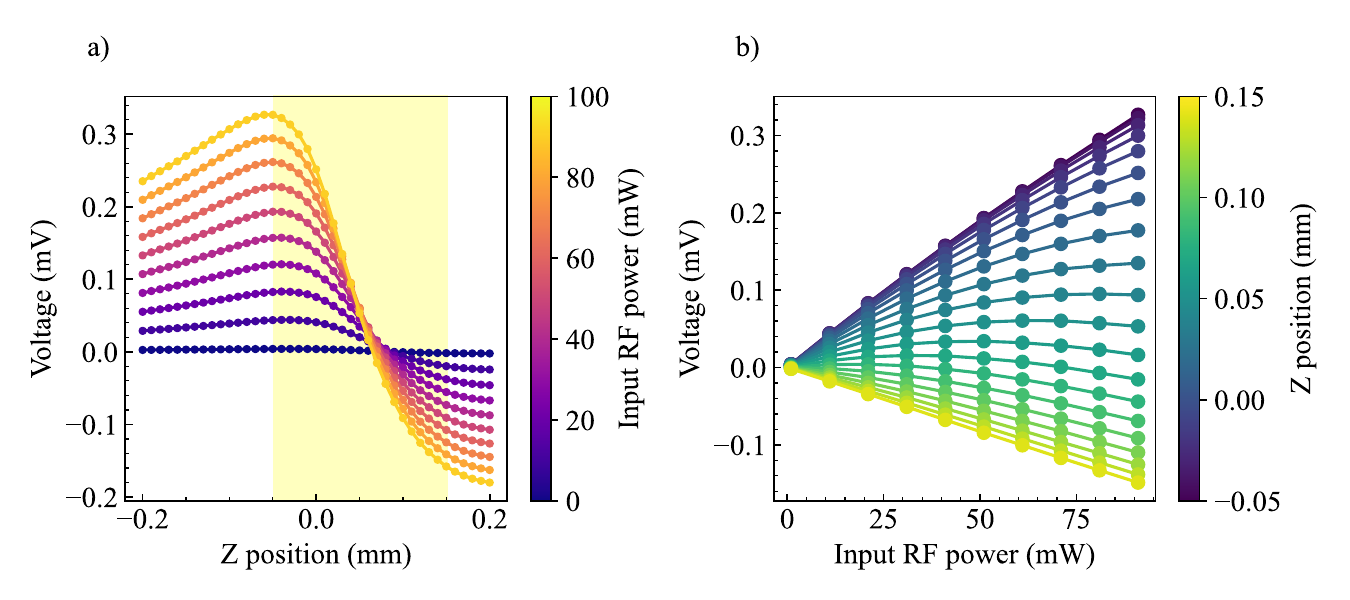}
    \captionof{figure}{(a) Experimental spin diode voltage versus the z position in the field gradient, (b) synaptic behavior.}
\end{minipage}
\noindent
\begin{minipage}{0.58\textwidth}
    \includegraphics[width=\textwidth]{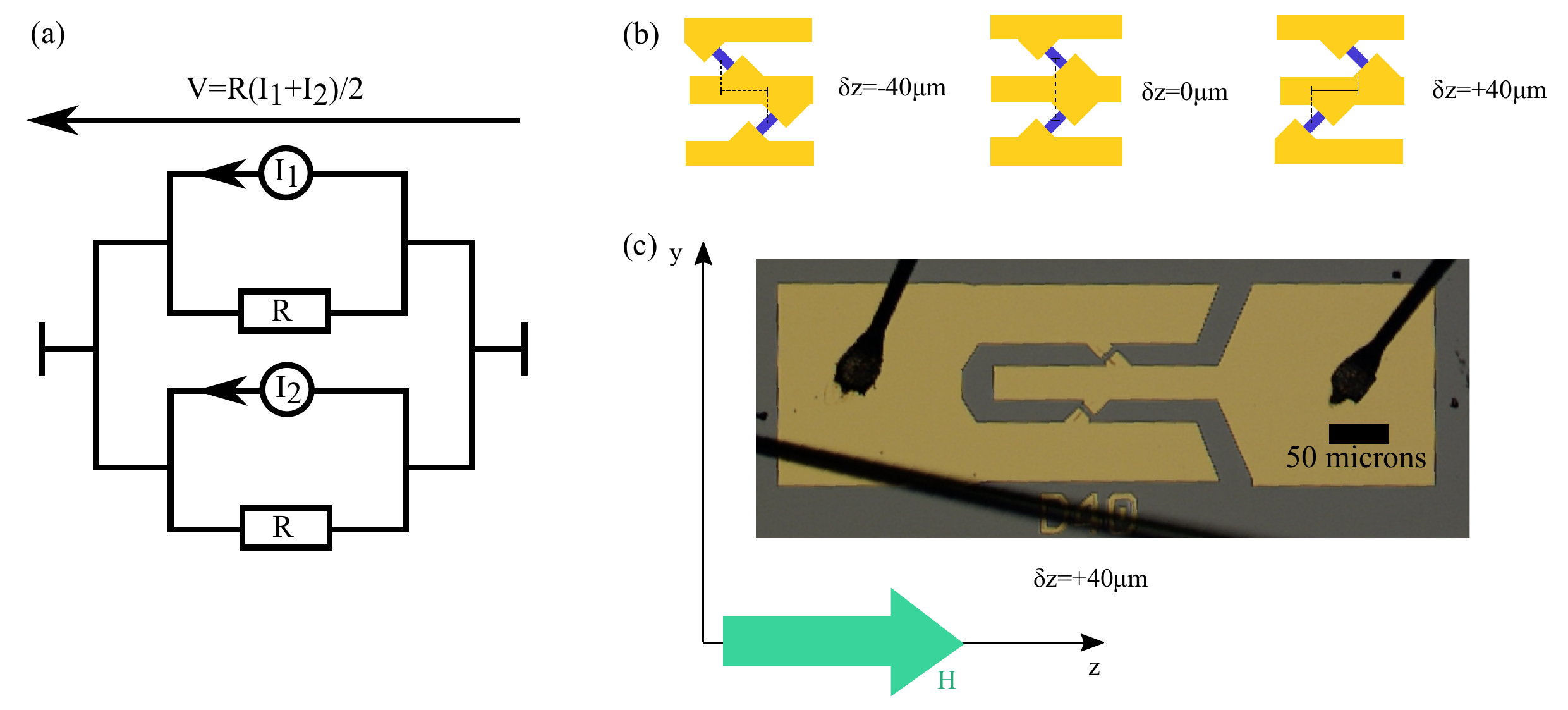}
    \captionof{figure}{Double diode synapse electric scheme (a), geometry for different $\delta z$ (b) and experimental realisation (c).}
\end{minipage}
\noindent
\begin{minipage}{0.4\textwidth}
    \includegraphics[width=\textwidth]{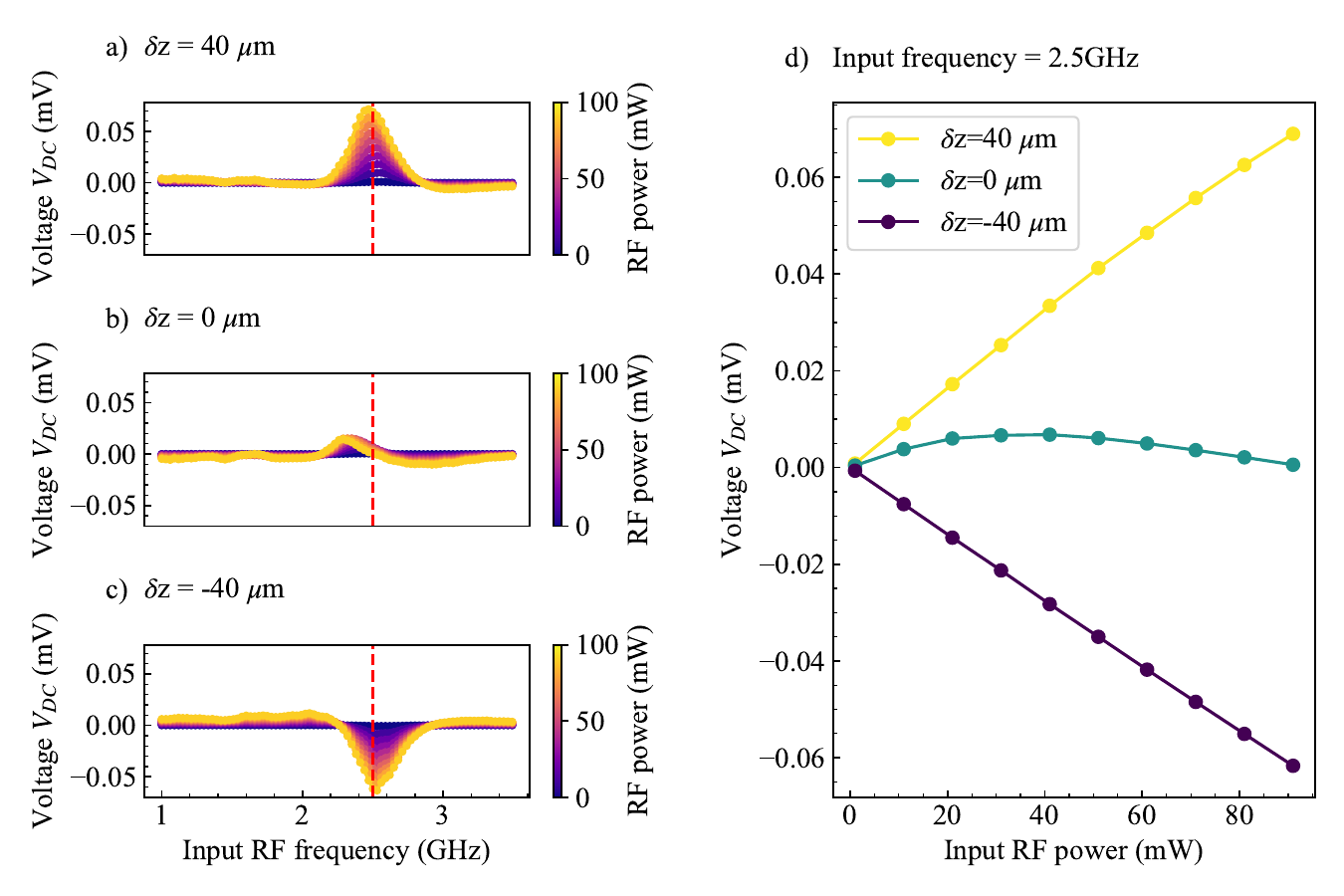}
    \captionof{figure}{Double diode synapse spin diode voltage versus frequency for different $\delta z$ (a)(b)(c) and associated synaptic behavior (d).}
\end{minipage}
\noindent
\begin{minipage}{0.48\textwidth}
    \includegraphics[width=\textwidth]{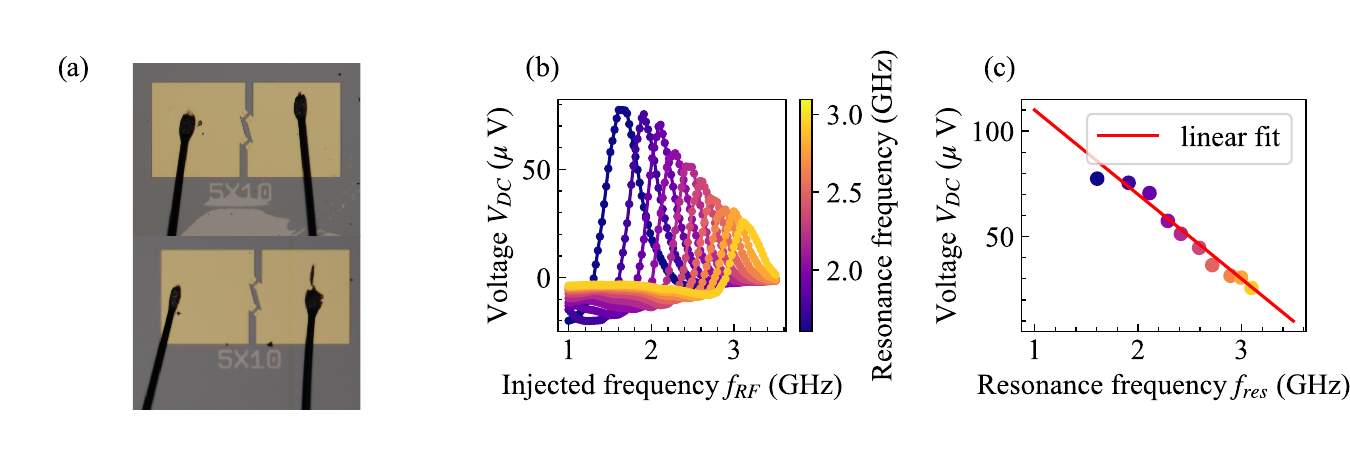}
    \captionof{figure}{(a) Calibration diode. (b) Spin diode voltage amplitude versus frequency, (c) linear fit of this dependance.}
\end{minipage}
\noindent
\begin{minipage}{0.48\textwidth}
    \includegraphics[width=\textwidth]{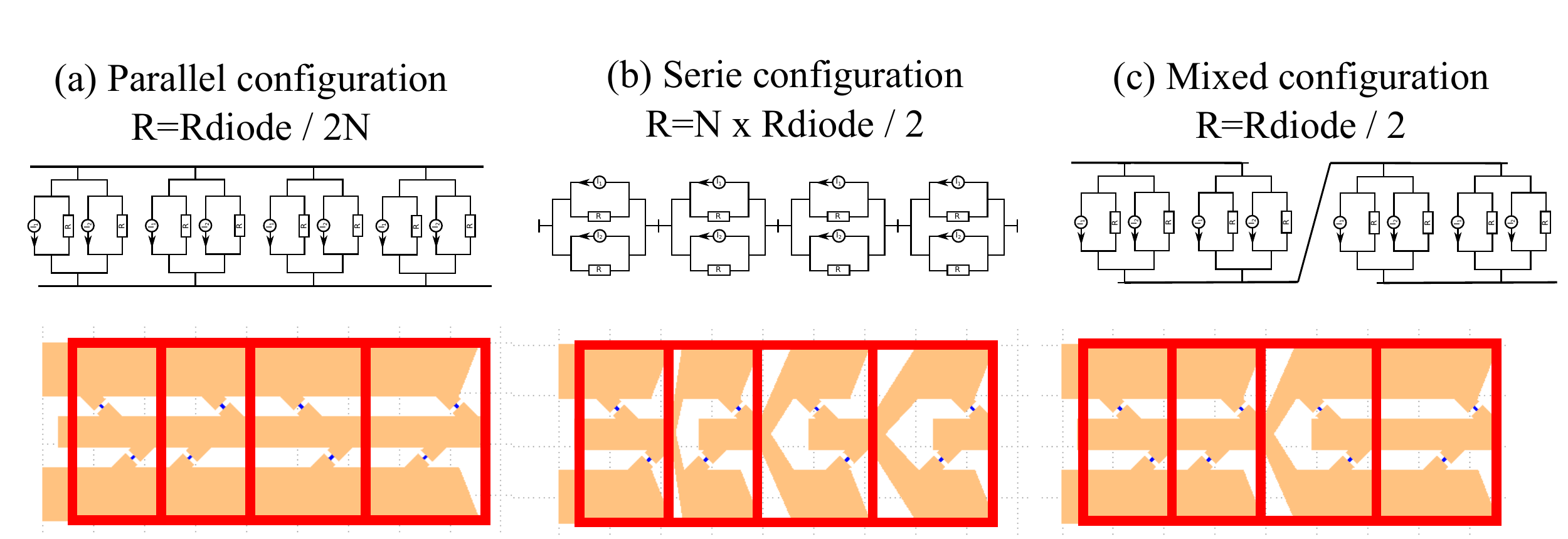}
    \captionof{figure}{Different synapse chaining, in parallel (a), series (b) and mixed (c).}
\end{minipage}
\noindent
\begin{minipage}{0.38\textwidth}
    \includegraphics[width=\textwidth]{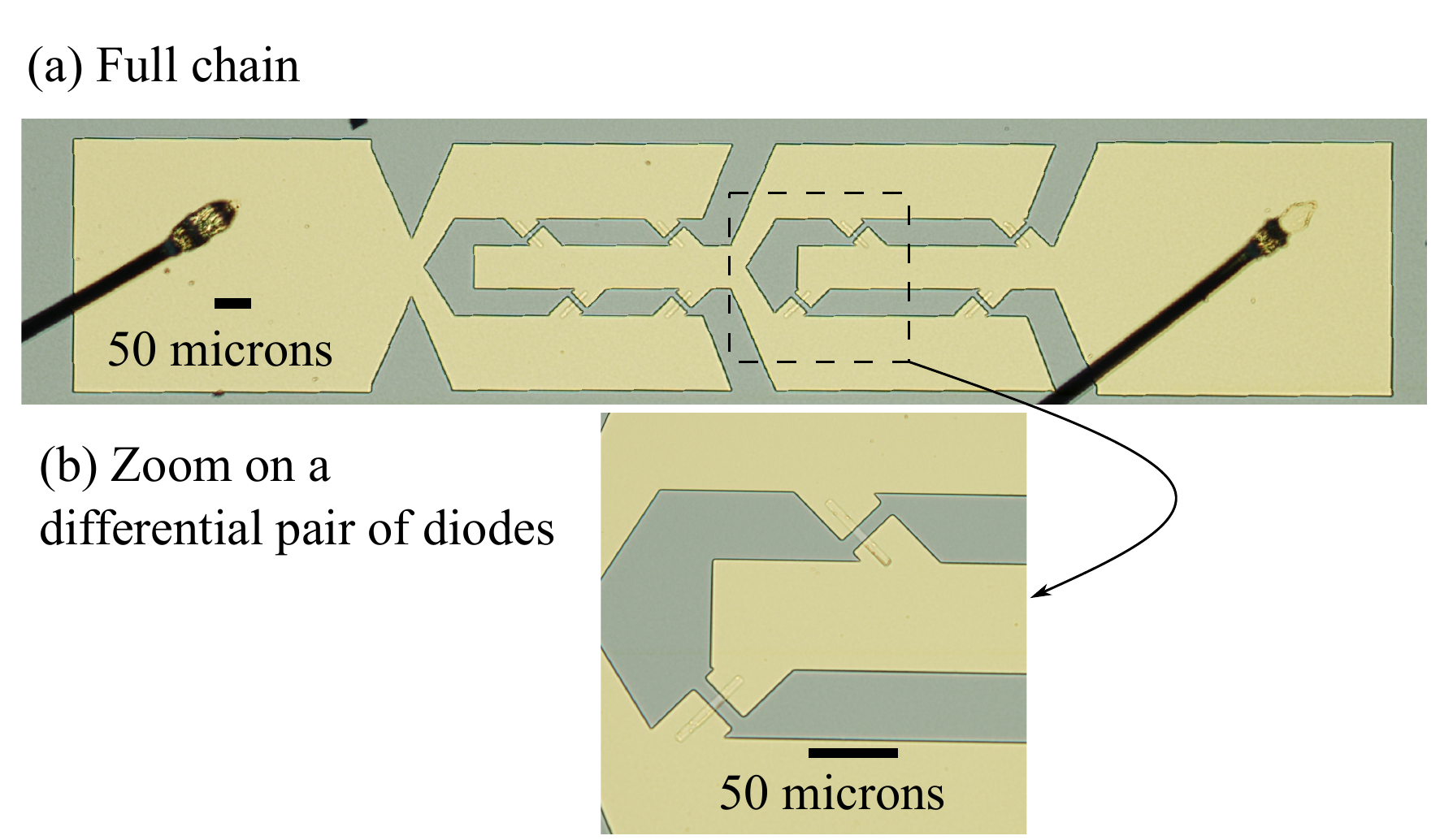}
    \captionof{figure}{Chain of 4 synapses after fabrication (a), and zoom on a synapse (b).}
\end{minipage}
\noindent
\begin{minipage}{0.6\textwidth}
    \includegraphics[width=\textwidth]{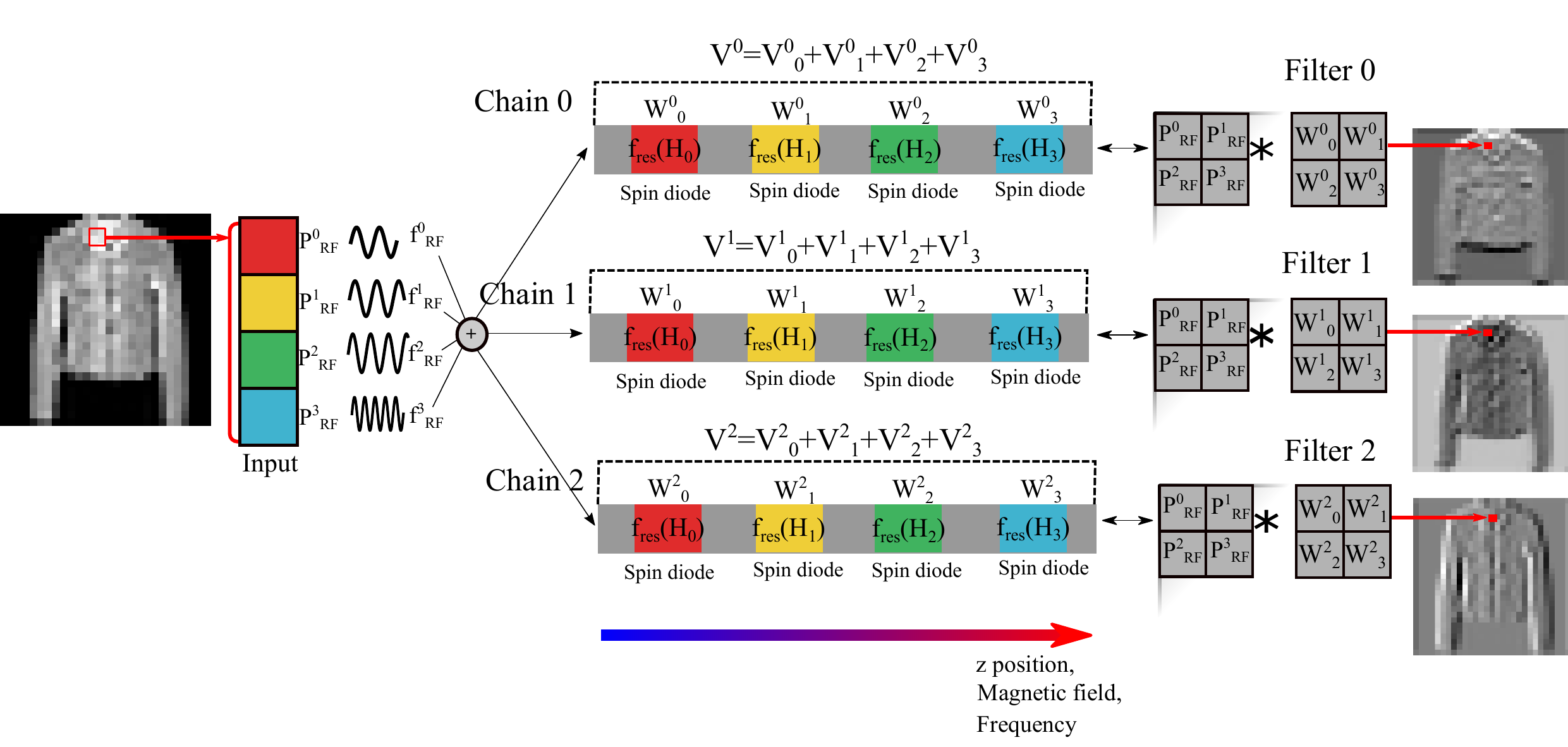}
    \captionof{figure}{Implementation of a convolutional layer with 3 chains of spintronic synapses.}
\end{minipage}
\noindent
\begin{minipage}{1\textwidth}
    \includegraphics[width=\textwidth]{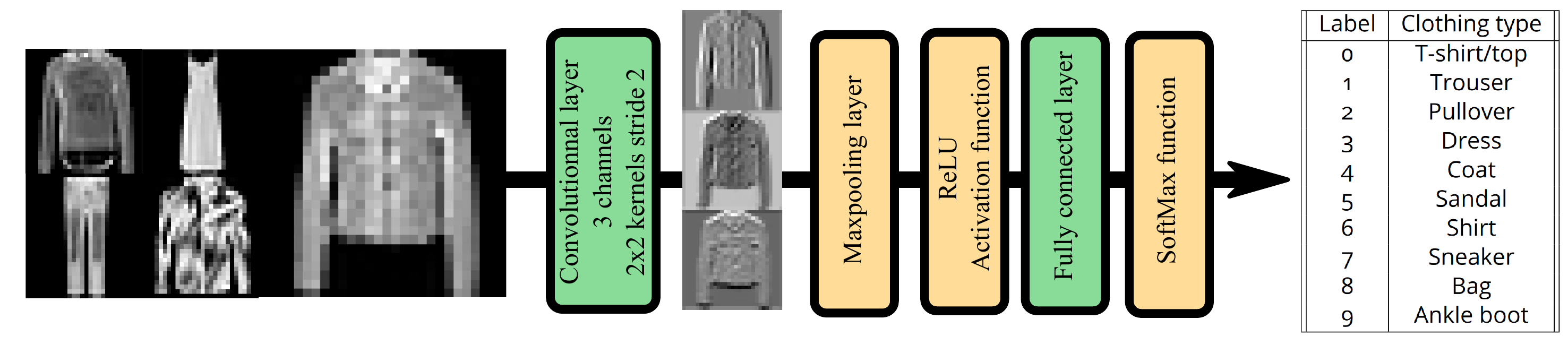}
    \captionof{figure}{Convolutional network for FashionMNIST classification.}
\end{minipage}
\begin{minipage}{0.3\textwidth}
\resizebox{1\textwidth}{!}{
\begin{tabular}{||c|c| c |c| c ||}
 \hline
  & $w_0$& $w_1$& $w_2$& $w_3$\\
 \hline

chain 0 &  & & & \\
 \hline
theoretical weights & -0.1322& -0.3949&-0.2745&  0.3621\\
experimental weights& -0.2684& -0.4309& -0.2949&  0.3289\\
$\delta z (\mu m)$& -43& -75& -47& 51\\
 \hline
chain 1 &  & & & \\
 \hline
theoretical weights & -0.2338&  0.2530&-0.2336&  0.2327\\
experimental weights& -0.2527 & 0.2153 &-0.141 & 0.1265\\
$\delta z (\mu m)$& -62& 64& -60& 62\\

 \hline
chain 2 &  & & & \\
 \hline
theoretical weights& -0.2177& -0.1121&0.2409&  0.1869\\
experimental weights& -0.2835& -0.1432&  0.2204  & 0.1573\\
$\delta z (\mu m)$& -63& -11& 82& 54\\
 \hline
\end{tabular}}
\captionof{figure}{Theoretical weights, associated $\delta z$ and obtained experimental weight values for the 3 chains.}
\end{minipage}
\noindent
\begin{minipage}{0.65\textwidth}
    \includegraphics[width=\textwidth]{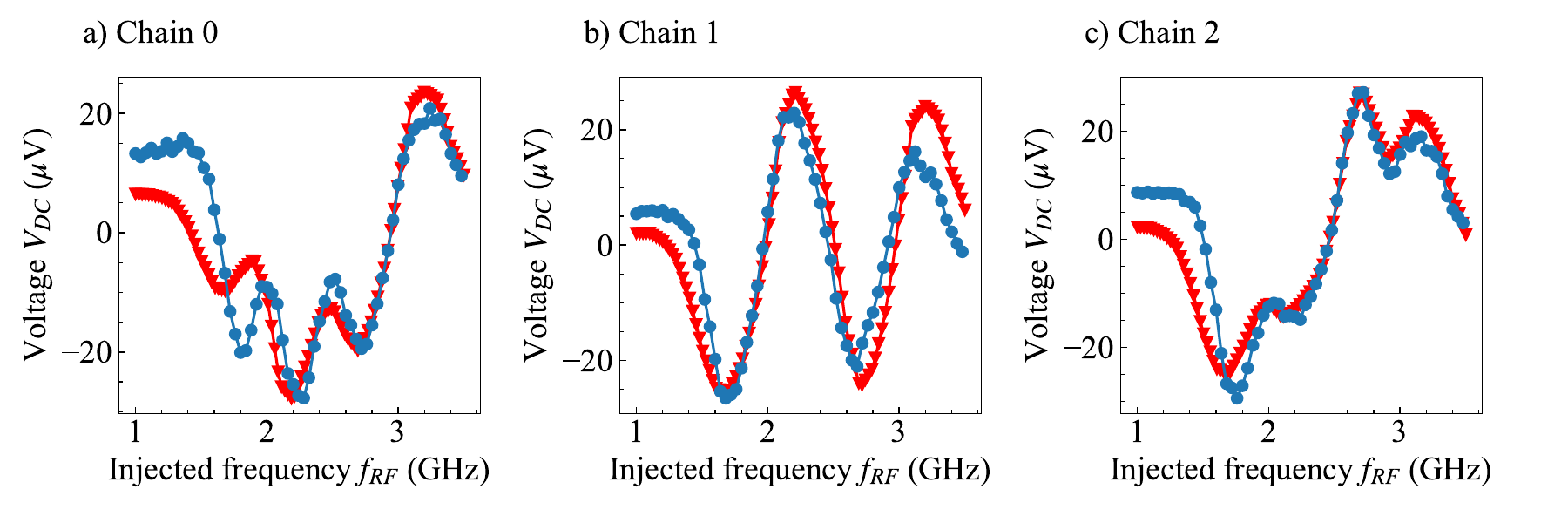}
    \captionof{figure}{Theoretical (red) and experimental (blue) spin diode voltages versus frequency for the 3 chains.}
\end{minipage}
\noindent
\begin{minipage}{0.28\textwidth}
    \includegraphics[width=\textwidth]{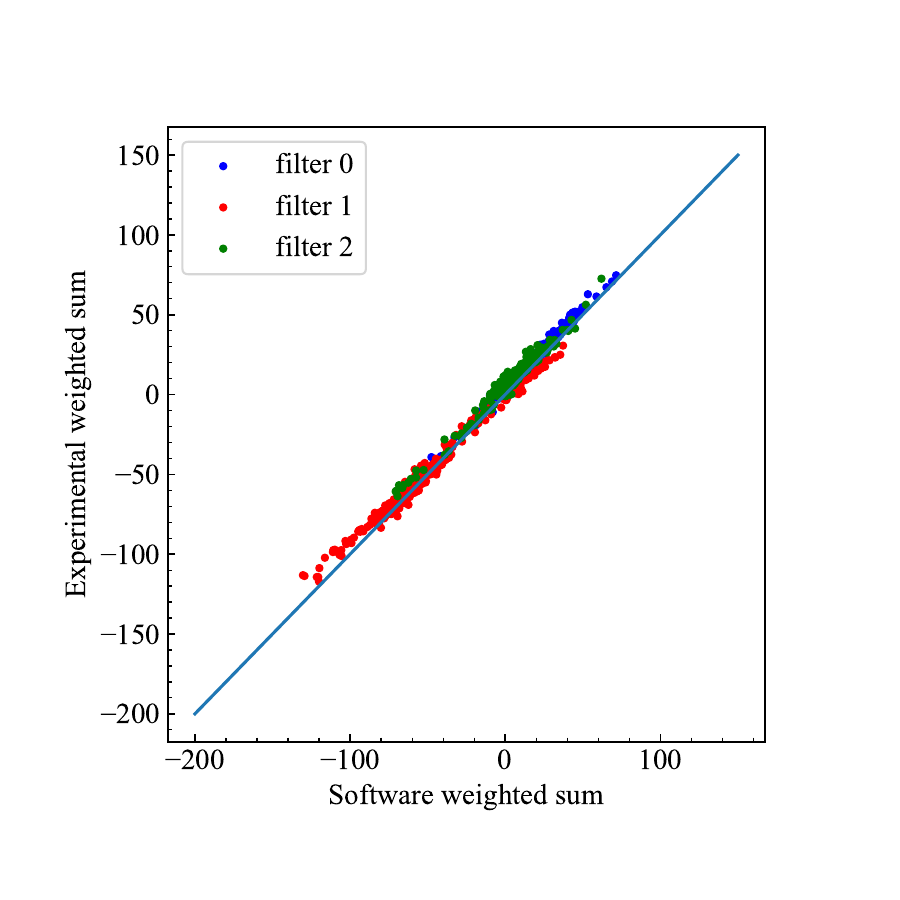}
    \captionof{figure}{Comparison between experimental and theoretical weighted sum, voltage values have been converted to pixel values.}
\end{minipage}
\noindent
\hfill
\begin{minipage}{0.38\textwidth}
    \includegraphics[width=\textwidth]{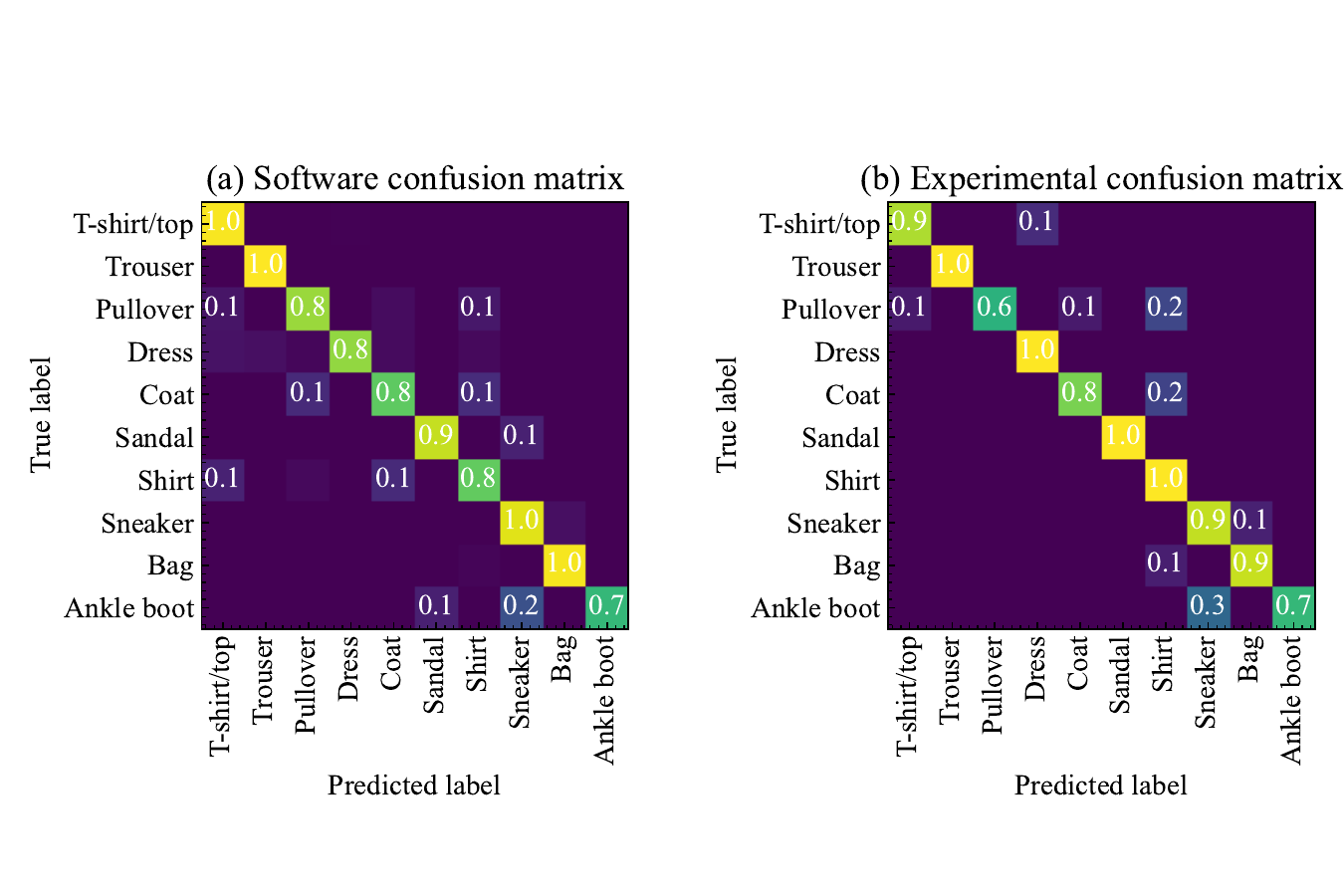}
    \captionof{figure}{Confusion matrices for the pure software network and the hybrid hardware-software network on the 100 first images of the dataset.}
\end{minipage}
\noindent
\hfill
\begin{minipage}{0.28\textwidth}
    \includegraphics[width=\textwidth]{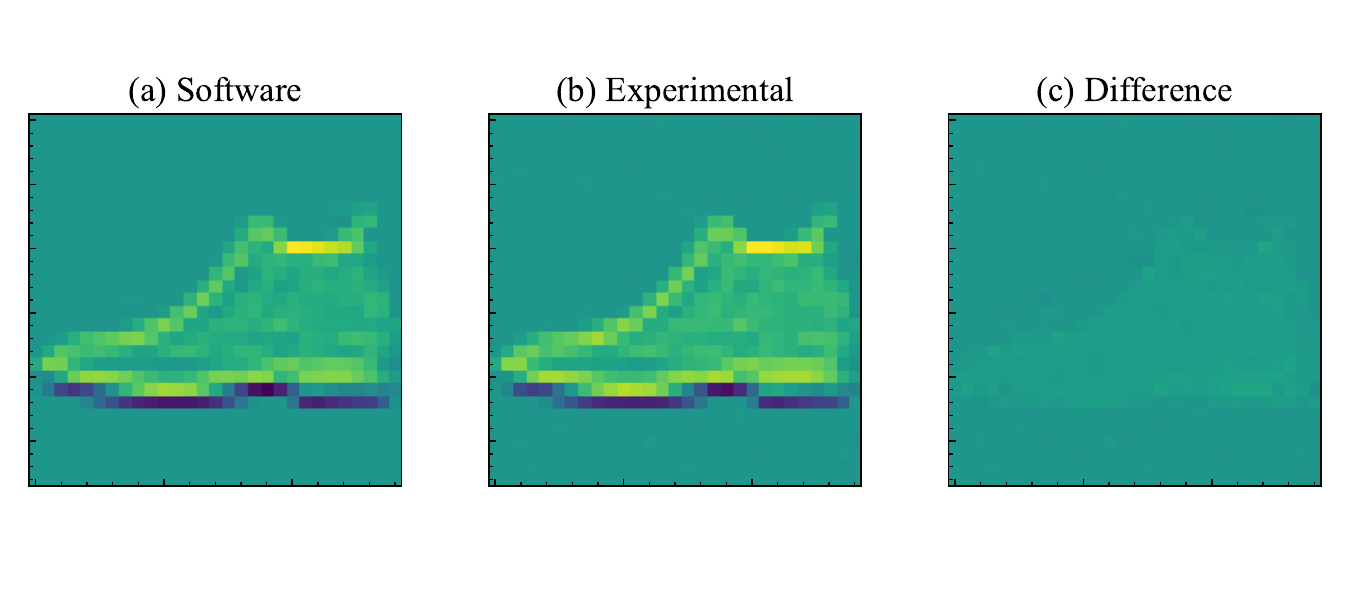}
    \captionof{figure}{Shoe image convoluted by the software network (a) and by the hardware-software network (b), (c) displays the difference between the two images.}
\end{minipage}

\end{document}